\documentclass[12pt,draftcls,onecolumn]{IEEEtran}

\usepackage{amssymb}
\usepackage{amsmath}
\usepackage[dvips]{graphicx}
\usepackage{cite}
\usepackage{citesort}

\usepackage{subcaption}

\newcommand{\imag}{{\rm i}}

\newcommand{\Ubf}{\boldsymbol{U}}
\newcommand{\Ibf}{\boldsymbol{I}}

\newcommand{\Pbf}{\boldsymbol{P}}
\newcommand{\Qbf}{\boldsymbol{Q}}
\newcommand{\cbf}{\boldsymbol{c}}

\newcommand{\xbf}{\boldsymbol{x}}
\newcommand{\ybf}{\boldsymbol{y}}

\newcommand{\Scal}{\mathcal{S}}

\newcommand{\Ccal}{\mathcal{C}} 
\newcommand{\Ucal}{\mathcal{U}} 
 
\newcommand{\Gcal}{\mathcal{G}}

\newcommand{\Ebb}{\mathbb{E}}

\newcommand{\grass}[3]{ {\Gcal}_{#2,#3}^{ \mathbb #1 }}

\allowdisplaybreaks[1]
\newcommand{\e}{{\rm e}}

\makeatletter

\newcommand{\Rmnum}[1]{\expandafter\@slowromancap\romannumeral #1@}

\newcommand{\dd}{\,\mathrm{d}}

\begin{document}
\title{Volume of Metric Balls in High-Dimensional Complex Grassmann Manifolds} 
\author{Renaud-Alexandre Pitaval, Lu Wei, Olav Tirkkonen, and Jukka Corander 
\thanks{Renaud-Alexandre Pitaval was with the Department of Mathematics and Systems Analysis, Aalto University, Finland (email: renaud-alexandre.pitaval@alumni.aalto.fi).
Lu Wei and Jukka Corander are with the Department of Mathematics and Statistics, University of Helsinki, Finland (email:\{lu.wei, jukka.corander\}@helsinki.fi). 
Olav Tirkkonen is with the Department of Communications and Networking, Aalto University, Finland (email: olav.tirkkonen@aalto.fi).

R.-A. Pitaval and L. Wei contributed equally to this work. 
 } 						
\thanks{Part of this work will be presented at the 2015 IEEE Information Theory Workshop in Korea.}
}

\maketitle

\begin{abstract} 
Volume of metric balls relates  to rate-distortion theory and  packing bounds on codes. 
In this paper, the volume of  balls in complex Grassmann manifolds is evaluated for an arbitrary radius. 
The ball is defined as a set of hyperplanes of a fixed dimension with reference to a center of possibly different dimension, 
and a generalized chordal distance for unequal dimensional subspaces is used. 
First, the volume is reduced to one-dimensional integral representation. 
The overall problem boils down to evaluating a determinant of a matrix of
the same size as the subspace dimensionality. Interpreting this determinant as a characteristic function of the Jacobi ensemble, an asymptotic analysis is carried out.  
The obtained asymptotic volume is moreover refined using moment-matching techniques to provide a tighter approximation in finite-size regimes. 
Lastly, the pertinence of the derived results is shown by rate-distortion analysis of source coding on Grassmann manifolds.
\end{abstract}

\section{Introduction}

A Grassmann manifold is the collection of subspaces of a given
dimension in a vector space. Grassmann manifolds find many applications due to their relation to  eigenspaces of matrices, see e.g. in~\cite{Conway1996,Edelman,YeLim}.
In the context of multi-antenna transmissions, complex Grassmannian codes have notably been used for non-coherent space-time coding~\cite{AgrawalTIT01,Tse,AshikhminTIT2010} and channel-aware precoding~\cite{LoveTIT03,Mukkavilli03}. 

A metric ball in the Grassmann manifold is analogous to a spherical cap on a sphere and evaluating its volume is critical for several performance measures of Grassmannian codes. 
When constructing codes as point-sets with the largest possible minimum distance, the volume of a metric ball directly applies to packing bounds on codes~\cite{Barg,henkel}. 
Moreover, rate-distortion theory on Grassmann manifolds has been extensively applied to channel quantization analysis of precoded MIMO systems~\cite{Roh,Gian06,Dai08,DaiOnOff}. In this source-coding context, the volume of metric ball is closely related to the cumulative distribution of quantization errors for a uniformly distributed source.

Computing the volume of metric ball in manifolds is recognized to be a difficult task. For the Grassmann manifold, this was addressed in several previous works. 
The volume of a metric ball in generic Grassmannians was derived for line packing with 
in~\cite{Mukkavilli03}, whereas asymptotic evaluations  for arbitrary subspace dimension were provided
in~\cite{Barg}, and in~\cite{DaiGrowth07} for balls with different dimensional center. A small ball approximation
was considered in~\cite{henkel} which was later derived exactly
in~\cite{Dai08} for balls of radius less than one. More recently, an exact volume formula for packing (2D) planes has been derived with application to massive MIMO~\cite{2014Zhu}. 
While the range of validity corresponding to the small ball volume has an exponential explosion in codesizes  with large dimension, the known asymptotics show slow convergences, providing only asymptotic scaling laws. 

The goal of this paper is to provide an accurate but simple volume approximation for regimes not covered by  previous works. 
The obtained asymptotics is complementary to the result in~\cite{Dai08} and with faster convergence than in~\cite{Barg,DaiGrowth07}; it is however limited to the complex case and its derivation is not straightforwardly generalizable to real Grassmannians. 
We consider the case of possibly unequal dimension
between elements in the ball and the center of the ball as
in~\cite{Dai08,DaiGrowth07}. Related problems with subspaces of
non-equal dimensions arise for example
in~\cite{JindalISIT06,AshikhminTIT2010,Thomas2014,LeeTSP14}. We start by
discussing generalizations of the well-known Grassmann chordal
distance~\cite{Conway1996} for subspaces of different dimensions, as well as relevant symmetries of the volume of ball with  our choice of distance. 
Then, from the known representation of the volume of ball as a multi-dimensional integration over principal angles~\cite{James,Barg,Adler,Dai08}, we reduce the problem to a one-dimensional integral related to a Fourier transform. The formulation is valid for any radius, and reduces the problem to the evaluation of a determinant that only depends of the dimension parameters. Accordingly, this integral can be computed exactly with fixed parameters and we provide several examples illustrating its versatility. The exact volumes have different polynomial representations in different ranges of integer part of the squared radius.  Radius less than one~\cite{Dai08} is itself a specific regime. From this, it can be anticipated that even though an exact generic formula could be derived, it will most probably be a linear combination of special functions similarly than in~\cite{2014Zhu}. Such representation may then not be very  amiable from application perspective, as for example one often  needs to invert the volume to computes bounds on codes. 

To provide a good and relevant approximation for large-dimensional Grassmann manifolds, an asymptotic analysis is further carried out.  
As all the dimension-related parameters are concentrated in a determinant inside the one-dimensional integral reformulation, the problem reduces to study the asymptotic behavior of the determinant.
This determinant is the partition function of the so-called time-dependent Jacobi ensemble. Interpreting it as a characteristic function of a linear spectral statistics, its asymptotic Gaussianity can be leveraged from random matrix theory~\cite{Johansson1997}. This leads to an asymptotic formulation of the volume of a metric ball in term Gauss error functions, which however appears loose  in some finite regimes.  We provide a finite-size correction to the asymptotic formula via the exact moments of the considered linear statistics, leading to a tighter volume approximation while preserving the simplicity of the asymptotic form. 

Finally, the derived asymptotic formula of the volume of metric ball (with its finite-size correction) is applied to rate-distortion theory on Grassmann manifolds.
With increasing dimensions, it is shown to provide a good estimate of the rate-distortion trade-off in a source coding problem. The correction is notable compared with using a small ball approximation outside of its regime of validity.   

The rest of this paper is organized as follows. Pertinent definitions and properties are given in Section II. The volume of metric ball is reduced in Section III to a single-fold integral leading to some examples of exact derivations. In Section IV, the asymptotic behavior of volume of metric ball is derived and corrected by moment-matching techniques for finite-size applications. In Section V, the derived expression is applied to source-coding on the Grassmann manifold. The paper is concluded in Section VI.

\section{Preliminaries}

\subsection{Grassmann Manifolds and Chordal Distance}

The complex Grassmann manifold $\grass{C}{n}{p}$ is the collection of
$p$-dimensional subspaces in an ambient $n$-dimensional complex
vector space $\mathbb{C}^{n}$. This is a homogeneous space of the
unitary group $\Ucal_n$ as there is a left action of $\Ucal_n$ on
$\grass{C}{n}{p}$ that acts transitively. A plane $P\in
\grass{C}{n}{p} $ can be described by infinitely many orthogonal bases
leading to non-unique semi-unitary matrix representation $\Pbf \in
\mathbb{C}^{n \times p}$, such that $\Pbf^\dag \Pbf = \Ibf_p$, where
$()^\dag$ is the conjugate transpose of a matrix.

There are several possible choices to define a distance on the
Grassmann manifold. We consider the chordal distance~\cite{Conway1996,Edelman} which is related to an embedding of the Grassmannian to an Euclidean sphere, and has been prominently used in
the literature~\cite{AgrawalTIT01,LoveTIT03,Love:limited,Dai08,Bachoc08}. The chordal distance is well-defined between subspaces with equal dimensions. However, one can find slight variations for its generalization to subspaces of unequal dimensions. We will use the same definition as e.g.~\cite{JindalISIT06,Dai08,LeeTSP14} arising from the concept of the principal angles;  discussions on its theoretical foundation can be found in the recent work~\cite{YeLim}. 

In~\cite{YeLim}, it is shown that any measure of distance that only
depends on the relative position between two subspaces must be a
function of the principal angles. The collection of principal angles
provides 
the relative position between subspaces which is transitive under the
action of the unitary group. However, compressing this ``vector-like
distance'' to a classical scalar distance $d$, one loses transitivity.
Grassmann manifolds are not in general two-point homogeneous
spaces~\cite{Wang52}: one cannot necessarily find a unitary mapping
between two pairs of equidistant points, i.e., a pair $(P,Q)$ cannot always
be mapped to a pair $(P',Q')$ even if $d(P,Q)=d(P',Q')$.

Consider two integers $p$ and $q$ satisfying $p , q\leq n$ and
$m=\min(p,q)$. Given $P\in \grass{C}{n}{p}$ and $Q\in \grass{C}{n}{q}$
with respective orthonormal bases $\Pbf \in P$, $\Qbf \in Q$ one can
define $m$ principal angles~\cite{James} between these two subspaces.
We denote the principal angles by $\theta_1 \ldots \theta_m \in [0,
  \frac{\pi}{2} ]$. 
They are independent of the choice of coordinates and can be computed
via the singular value decomposition of $\Pbf^\dag \Qbf $ 
whose 
singular values are $\{\cos \theta_i\}_{i=1}^{m}$. 
The considered square chordal distance is given by
\begin{eqnarray} 
d^2_c(P,Q) &=& \sum_{k=1}^{\min(p,q)} \sin^2( \theta_k) \nonumber\\
 &=&\min(p,q)- \sum_{k=1}^{\min(p,q)} \cos^2( \theta_k) \nonumber\\
 &=&\min(p,q) - \| \Pbf^\dag \Qbf \|_F^2 \label{eq:dc}   .
\end{eqnarray}

\subsection{Relation to Other Chordal Distances}

It is noted in~\cite{YeLim} that $d_c$ gives a notion of distance in
a sense of  a distance from a point to a set, but it does not
give a metric 
between subspaces of different dimensions 
since two distinct subspaces of different dimensions can have distance
zero. Nevertheless, a simple variation 
leading to a properly-defined metric function is given
in~\cite{YeLim}. This is obtained by assigning $|p-q|$ additional
principal angles with value $\frac{\pi}{2}$ for the dimensions
mismatched between $P$ and $Q$. One can then define
\begin{eqnarray} 
d^2_{c\#}(P,Q) &=& \max(p,q) - \sum_{k=1}^{\min(p,q)} \cos^2( \theta_k).
\end{eqnarray} 

The chordal distance has also been generalized to subspaces of
different dimensions from their corresponding projection operators
in~\cite{AshikhminTIT2010}. This corresponds to the Euclidean distance
of a spherical embedding into $\Scal^{n^2-1}({\scriptstyle
  \frac{\sqrt{n}}{2}})$, each Grassmannian being itself embedded in a
different cross-sectional sphere, specifically 
the $p$-dimensional subspaces to
$\Scal^{n^2-2}({\scriptstyle\sqrt{\frac{p(n-p)}{n}}} )$ and
the $q$-dimensional subspaces to
$\Scal^{n^2-2}({\scriptstyle\sqrt{\frac{q(n-q)}{n}}})$~\cite{Conway1996,henkel}.
This gives a proper metric which can be 
expressed in term of principal angles as
\begin{eqnarray} 
d^2_{c*}(P,Q) &=& \| \Pbf \Pbf^\dag - \Qbf \Qbf^\dag  \|_F^2 \nonumber \\
&=& p+q - 2 \sum_{k=1}^{\min(p,q)} \cos^2( \theta_k).
\end{eqnarray} 

Finally, we suggest a third  metric for subspaces of unequal dimensions which
provides a slight reduction in the dimension of the embedding. The
main observation is that all Grassmannians  in $n$ dimensions can be embedded in a single
sphere $\Scal^{n^2-2}$. This holds in fact for any flag manifold as
well~\cite{PitavalITW13}. To obtain this, one must detrace the
projectors and rescale them as $\tilde{P} =
\sqrt{\frac{n}{p(n-p)}}(\Pbf \Pbf^\dag - \frac{p}{n} \Ibf) $ and
$\tilde{Q} = \sqrt{\frac{n}{q(n-q)}}(\Qbf \Qbf^\dag - \frac{q}{n}
\Ibf) $, then $\tilde{P}$ and $\tilde{Q}$ lie on the same unit sphere
$\Scal^{n^2-2}$. The corresponding Euclidean distance is
\begin{eqnarray} 
d^2_{c \star}(P,Q) &=& \| \tilde{P} - \tilde{Q} \|_F^2 \nonumber\\
 &=&   K_1 - K_2 \| \Pbf^\dag \Qbf \|_F^2  \nonumber\\
										&=&  K_1-  K_2 \sum_{k=1}^{\min(p,q)} \cos^2( \theta_k)
\end{eqnarray} 
with $ K_1=  2+2 \sqrt{\frac{p q}{(n-p) (n-q)}}$ and $ K_2= \frac{2 n}{\sqrt{pq(n-p)(n-q)}}$.

All the distances $d_{c}$, $d_{c \#}$, $\frac{1}{\sqrt{2}} d_{c *}$
and $\frac{1}{\sqrt{K_2}} d_{c \star}$ reduce to the classical
definition of chordal distance for $p=q$. In the rest of this paper,
we will keep the definition of the distance $d_c$ as
in~\eqref{eq:dc} due to its compactness in term of principal
angles and for consistency with~\cite{JindalISIT06,Dai08,LeeTSP14}. Corresponding results
can be easily extended to the other distances considered above since
$d_c$ includes the main information of interest, and differs only by
constant factors from the  other metrics.

\subsection{Metric Ball and Normalized Volume}
Define the metric balls of $q$-dimensional subspaces with distance at most $r$ from the  $p$-dimensional center  $P \in \grass{C}{n}{p}$ by 
\begin{equation} 
B_{\scriptscriptstyle P,q}(r) = \left\{ Q \in \grass{C}{n}{q} \; : \quad   d_c(P,Q) \leq r \right\} .
\end{equation} 
The ball $B_{\scriptscriptstyle P,q}(r)$ is a subset of
$\grass{C}{n}{q}$ though it is defined with reference to a point in
$\grass{C}{n}{p}$.

We consider the invariant Haar measure $\mu$, defining an
uniform distribution on $\grass{C}{n}{q}$. For any measurable set
$\mathcal{S} \subset \grass{C}{n}{p}$ and any $\Ubf \in \Ucal_n$, the
Haar measure satisfies
\begin{equation} \mu(\Ubf \mathcal{S} ) = \mu(\mathcal{S} ). \end{equation}
The quantity $\mu (B_{\scriptscriptstyle P,q}(r))$ is independent of
the center $P$, and we will simply write $\mu(B_{\scriptscriptstyle
  p,q}(r) )$ or even $\mu(B(r) )$ when there is no ambiguity.

The invariant measure can be interpreted as a normalized
volume 
\begin{equation}\mu (B_{\scriptscriptstyle p,q}(r))= \frac{{\rm
    vol}(B_{\scriptscriptstyle p,q}(r))}{{\rm vol}(\grass{C}{n}{q})}
		\end{equation} 
where with our choice of distance the corresponding volume of the
Grassmann manifold is~\cite{PitavalAsilomarBall}
\begin{equation}
{\rm vol}(\grass{C}{n}{q}) = \pi^{q(n-q)} \prod_{i=1}^q \frac{(q-i)!}{(n-i)!}.
\end{equation}

It is shown in~\cite{YeLim} that distances from principal angles as
considered here are independent of the dimension $n$ of the ambient
space. The choice of ambient space, 
however, 
has an impact on the maximum possible value of $d_c$ and thus the
range of $\mu$.
From~\cite[Lem.~2]{Dai08} it can be deduced that the chordal distance
must satisfy $d_c(P,Q) \leq d_{\max}$ with
$d^2_{\max}=\min(p,q,n-p,n-q)$. As a consequence
$\mu(B_{\scriptscriptstyle p,q}(r) )$ is defined on the range $[0,\,
  d_{\max}]$, its maximum is $\mu(B_{\scriptscriptstyle
  p,q}(d_{\max}))=1 $, and the volume depends on $n$.
The dependence on the choice of ambient space can be understood by
considering e.g. the problem of packing 2D real planes. It is
impossible to find two fully orthogonal planes in $\mathbb{R}^3$ (i.e.
they intersect only in the zero vector), while it is possible in
$\mathbb{R}^4$.

\subsection{Symmetries and Complementary Balls}

Without loss of generality we will assume all along the paper that $p\leq q <n$ and $p+q \leq n$,
implying that $p\leq n/2$. Results in other parameter ranges can be reproduced using the chordal distance and the canonical isomorphism $\grass{C}{n}{p}
\cong \grass{C}{n}{n-p}$:
\begin{eqnarray}
\mu (B_{\scriptscriptstyle p,q}(r)) &=& \mu (B_{\scriptscriptstyle q,p}(r)) \label{eq:sym1}\\
\mu( B_{\scriptscriptstyle p,q}(r) ) &=& \mu(B_{\scriptscriptstyle n-p,n-q}(r))  \label{eq:sym2}.
\end{eqnarray}
These symmetries were used for volume computations in~\cite{Dai08} as for $p+q \geq n$, one can evaluate $\mu(B_{\scriptscriptstyle p',q'}(r))$ with $p'=n-p$  and $q'=n-q$ satisfying $p'+q' \leq n$.

An additional symmetry than can be used for extending results from one Grassmannian to another with  a different another range of radius values is as follows. 
Let $p,q$ satisfy $p\leq q <n$ and $p+q \leq n$. Then we have
\begin{equation} \label{eq:sym3}
\mu \left( B_{\scriptscriptstyle p,q}(r) \right) = 1-\mu \left( B_{\scriptscriptstyle p,n-q}\left(\sqrt{p-r^2} \right) \right)  ~. 
\end{equation} 

\begin{figure}[t]
\begin{center}
\includegraphics[width=4.5in]{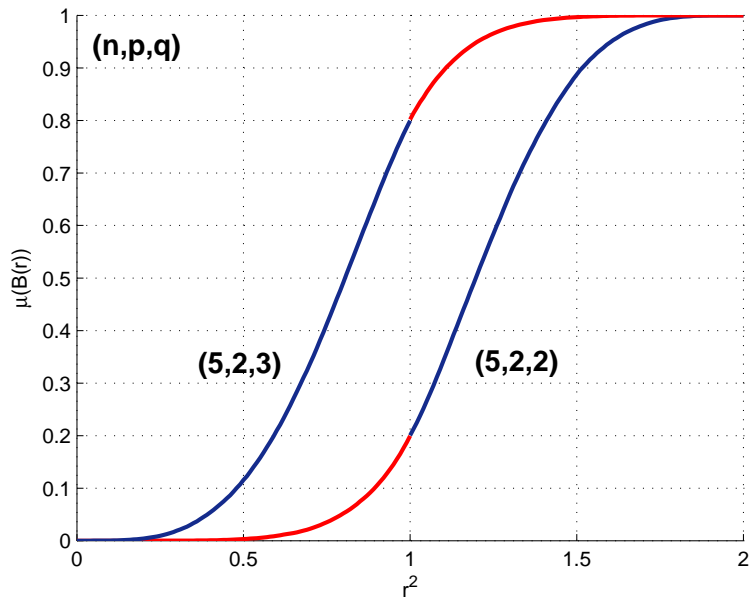} 
\end{center}
\vspace{-0.5cm}
\caption{Illustration of the symmetry~\eqref{eq:sym3}  
  between $\mu( B_{\scriptscriptstyle p,q}(r) )$ and $\mu(
  B_{\scriptscriptstyle p,n-q}(r) )$: one curve is 
the rotation of the other by $180^\circ $ around the median.  
\label{fig:symmetry}}
\end{figure}

The proof is in Appendix~\ref{app:sym3}. For the specific case $q=\frac{n}{2}$, one sees that the volume is a symmetric function in $r=\sqrt{p/2}$ since  $\mu( B(r) ) = 1-\mu( B(\sqrt{p-r^2}) )$.  
Combining~\eqref{eq:sym3} and the result of~\cite{Dai08}  directly leads to the following elementary exact evaluation of volume of balls for any radius with $n=4,\, p=q=2$:  
\begin{equation}
\label{eq:vol_G42}
\mu( B(r)) = \left\{	\begin{array}{ll}
								  \frac{1}{2}r^8 & \text{for } r \leq 1\\
									1-\frac{1}{2}(2-r^2)^4  & \text{for } r \geq 1
									\end{array} \right. .
\end{equation}
The symmetry~\eqref{eq:sym3} 
is illustrated in Figure~\ref{fig:symmetry}. The exact evaluation for $p=q$ and $r\leq 1$ in~\cite{Dai08} is highlighted in red. It can be directly used for computing the volume for $q=n-p$ and $r \geq \sqrt{p-1}$.

\subsection{Sphere-Covering/Packing Bounds}

When subspaces have the same dimension  $p=q$, a direct application of the volume of metric ball occurs in the evaluation of fundamental coding bounds.  
The Gilbert-Varshamov and Hamming bounds, derived from a sphere-covering and sphere-packing arguments, respectively, relate the code's cardinality to its minimum distance.
Namely, for any distance $\delta$, there exists a code $\Ccal\subset \grass{C}{n}{p}$ with cardinality $|\Ccal|$ such that 
\begin{equation}
\frac{1}{\mu(B(\delta))} \leq  |\Ccal|  
\end{equation}
while for any $(|\Ccal|,\delta)$-code $\Ccal \subset \grass{C}{n}{p}$ one must have 
\begin{equation}
|\Ccal|  \leq  \frac{1}{\mu(B(\frac{\delta}{2}))}. 
\end{equation}

\section{Exact Integral Formulations}
The volume of a metric ball in the Grassmann manifold is known to be expressible as a multivariate integration over principal angles. Here we reduce
the problem to a one-dimensional integral related to a Fourier transform.  
It is assumed without loss of generality that  $p\leq q <n$ and $p+q \leq n$ and complementary cases can be treated by symmetry. It is worth noting that as a consequence the range of radii of balls is $r\in[0,\sqrt{p}]$. 

\subsection{Multi-dimensional Integration}
An integration of the volume element on the Grassmann manifold can be split in three parts including two densities on Stiefel manifolds that can be fully integrated. The overall calculation reduces then to an integral over the marginal distribution of the  principal angles~\cite{James,Barg}.  
With the cosines of the principal angles $c_{i} = \cos \theta_i$,
$i=1 \ldots p$, the volume of a metric ball $\mu\left(B\left(r\right)\right)$ in complex Grassmann
manifolds can be written as a $p$-dimensional integral of the form~\cite{James,Adler,Dai08}   
\begin{equation}\label{eq:dv}
\mu\left(B\left(r\right)\right) = v_{n,p,q} \mkern-36mu  \int \limits_{\substack{0\leq c_{i}\leq1, \\ \sum_{i=1}^{p}(1-c_{i}^2)\leq r^2}}  \mkern-38mu \Delta^{2}(\cbf^2)\prod_{j=1}^{p}c_{j}^{2(q-p)}\left(1-c_{j}^2\right)^{n-p-q}\dd c_{i}^2,
\end{equation}
where the normalization constant $v_{n,p,q}$ is given by
\begin{equation}
v_{n,p,q} = \prod_{j=1}^{p} \frac{\Gamma\left( n-j+1\right)}{\Gamma\left(j+1\right)\Gamma\left(n-q-j+1\right)\Gamma\left(q-j+1\right)},
\end{equation}
and
\begin{equation}
\Delta(\cbf)=\det\left(c_{j}^{2(i-1)}\right)=\prod_{1\leq i<j\leq p}\left(c^2_{i}-c^2_{j}\right) \label{eq:vandDet}
\end{equation}
denotes a Vandermonde determinant. 
We note here that the volume element is unique up to a scaling factor (which is included into the overall normalization) and the choice of distance affects only the domain of integration.

Applying the change of variables $x_j =1- c_j^2$, Eq.~\eqref{eq:dv} simplifies to
\begin{equation}\label{eq:dv2}
\mu\left(B\left(r\right)\right) = v_{n,p,q} \mkern-16mu \int \limits_{\substack{0\leq x_{i}\leq1, \\ \sum_{i=1}^{p}x_{i}\leq r^2}} \mkern-16mu \Delta^{2}(\xbf)\prod_{j=1}^{p}x_{j}^{n-p-q}\left(1-x_{j}\right)^{q-p}\dd x_{j}
\end{equation}
where  $\Delta(\xbf) = \det\left(x_{j}^{(i-1)}\right)=\prod_{1\leq i<j\leq p}\left(x_{i}-x_{j}\right) $ similarly as in~\eqref{eq:vandDet}. 
The constraint \mbox{$\sum_{i=1}^{p}x_{i}\leq r^2$}  in the integral~(\ref{eq:dv2}) presents the main challenge to obtain an explicit expression. 

\subsection{Time-dependent Jacobi Ensemble}

To address the issue raised above, we rewrite the integral~\eqref{eq:dv2} by using an indicator function: 
\begin{equation}
\label{eq:vde}
\mu\left(B\left(r\right)\right)=  v_{n,p,q} 
\int \limits_{0}^{r^2}\int \limits_{0\leq x_{i}\leq1} \mkern-16mu \delta\Big(t-\sum_{j=1}^{p}x_{j}\Big)\Delta^{2}(\xbf)\prod_{j=1}^{p}x_{j}^{n-p-q}\left(1-x_{j}\right)^{q-p}\dd x_{j}\dd t.
\end{equation}
Here $\delta(\cdot)$ is Dirac delta function, which admits the following Fourier representation
\begin{equation}
\delta(t-a)=\frac{1}{2\pi}\int_{-\infty}^{\infty}\e^{\imag(t-a)\nu}\dd\nu. \label{eq:Diracfourier}
\end{equation}
Note that a similar idea of using an indicator function was considered in~\cite{Han} for evaluating volumes in the unitary group. 

Inserting~\eqref{eq:Diracfourier} into~(\ref{eq:vde}) and performing the integration over $t$, we arrive at
\begin{equation}
\label{eq:dvre}
\mu\left(B\left(r\right)\right) =\frac{1}{2\pi} \int_{-\infty}^{\infty}\frac{\imag}{\nu}\left(1-\e^{\imag r^2\nu}\right) D_{p}(\nu) \dd\nu,
\end{equation}
where 
\begin{equation}\label{eq:Dp}
D_{p}(\nu)=v_{n,p,q} \int\dots\int_{0\leq x_{j}\leq1}\Delta^{2}(\xbf)\prod_{j=1}^{p}x_{j}^{n-p-q}\left(1-x_{j}\right)^{q-p}\e^{-\imag\nu x_{j}}\dd x_{j}.
\end{equation}
Comparing~(\ref{eq:dv2}) to~(\ref{eq:dvre}) and~(\ref{eq:Dp}), we see that the reformulation amounts to eliminating the constraint $\sum_{j=1}^{p}x_{j}\leq r^2$ at the expense of introducing a deformation $\e^{-\imag\nu x_{j}}$ in the $p$-dimensional integral~(\ref{eq:Dp}). As such the main difficulty is now concentrated in evaluating $D_{p}(\nu)$  which is independent of the radius and only depends on dimension parameters. 
The integral in~(\ref{eq:Dp}) 
is  the partition function of the so-called time-dependent Jacobi ensemble~\cite{2010BCE}, which is the classical Jacobi ensemble deformed by $\e^{-\imag\nu \sum x_{j}}$.

\subsection{One-dimensional Integral Formula and Exact Evaluations}
It is possible to further simplify the volume formula to a one-dimensional integral. 
To proceed, we use the Andr\'{e}ief integral
identity~\cite{1883Andreief,Chiani}, see Appendix~\ref{a:Andreief}, as well
as the symmetry of the integrand to simplify the integral to
\begin{eqnarray}
D_{p}(\nu)&\!\!=\!\!& v_{n,p,q} \int_{0\leq x_{i}\leq1}\Delta^{2}(\xbf)\prod_{j=1}^{p}x_{j}^{n-p-q}\left(1-x_{j}\right)^{q-p}\e^{-\imag\nu x_{j}}\dd x_{j}\label{eq:dj}\\
&\!\!=\!\!&p! v_{n,p,q}\det\left(\int_{0}^{1}x^{i+j-2+n-p-q}(1-x)^{q-p}\e^{-\imag\nu x}\dd x\right) \label{eq:AndriefStep} \\
&\!\!=\!\!&p! v_{n,p,q}\det\left(B\left(\alpha,\beta\right)~_{1}F_{1}\left(\alpha,\alpha+\beta;-\imag\nu\right)\right),
\end{eqnarray}
where the last equality is obtained by~\cite[Eq.~3.383]{2007GR} with
\begin{eqnarray}
\alpha&=&i+j+n-p-q-1,\\
\beta&=&q-p+1.
\end{eqnarray}
Here
\begin{equation}
B\left(\alpha,\beta\right)=\frac{\Gamma(\alpha)\Gamma(\beta)}{\Gamma(\alpha+\beta)}
\end{equation}
is the Beta function and
\begin{equation}
~_{1}F_{1}\left(\alpha,\beta;x\right)=\sum_{k=0}^{\infty}\frac{(\alpha)_{k}}{(\beta)_{k}k!}x^{k}
\end{equation}
defines the hypergeometric function, where $(\alpha)_{k}=\Gamma(\alpha+k)/\Gamma(\alpha)$ is Pochhammer symbol.
Note that the step from~\eqref{eq:dj} to~\eqref{eq:AndriefStep} using 
Andr\'{e}ief identity could not be generalized to compute volumes in real Grassmann manifolds. This is because the Vandermonde determinant in~\eqref{eq:dj} is not squared in the real case~\cite{Dai08}.

Putting everything together, we obtain an integral representation of $\mu\left(B\left(r\right)\right)$ for any radius,
\begin{equation}
\label{eq:main}
\mu\left(B\left(r\right)\right)=\frac{p!v_{n,p,q}}{2\pi} 
\int_{-\infty}^{\infty}\frac{\imag}{\nu}\left(1-\e^{\imag r^2\nu}\right)\det\left(B\left(\alpha,\beta\right)~_{1}F_{1}\left(\alpha,\alpha+\beta;-\imag\nu\right)\right) \dd\nu.
\end{equation}

For the special case $p=q=n/2$, 
the above general result simplifies further to

\begin{equation}
\label{eq:qpn}
\mu\left(B\left(r\right)\right)=\frac{p!v_{n,p,q}}{2\pi}
\int \limits_{-\infty}^{\infty}\frac{\e^{\imag r^2\nu}-1}{(\imag\nu)^{p^{2}+1}}\det\left(\Gamma(i+j-1)\Big(1-\e^{-\imag\nu} \!\!\!\sum_{k=0}^{i+j-2}\frac{\left(\imag\nu\right)^{k}}{k!}\Big)\right)\dd\nu. 
\end{equation}

\begin{figure}[t]
\begin{center}
\includegraphics[width=4.5in]{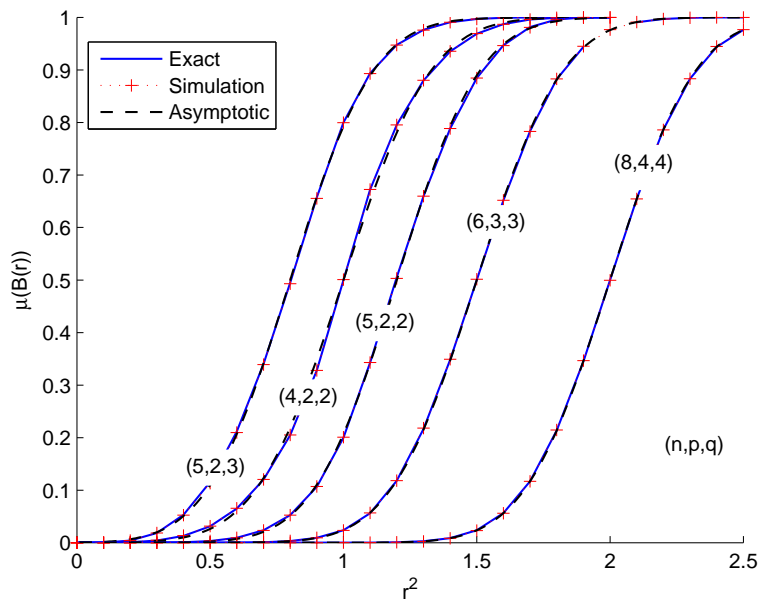} 
\end{center}
\vspace{-0.5cm}
\caption{Volume formulas~\eqref{eq:main} and~\eqref{eq:qpn} (Appendix~\ref{a:exemples}) for $\mu(B(r))$ versus simulations and asymptotic approximation~\eqref{eq:ve}.  \label{fig:simu}}
\end{figure}

The main technical difficulty in the single-integral formulation~\eqref{eq:main} and~\eqref{eq:qpn} lies in computing a determinant of a $p \times p$ matrix. This is tractable and can be further carried out for specific small values of the parameters $n,\,p$ and $q$. We list some of the results in Appendix~\ref{a:exemples}. 
The obtained expressions are verified to match Monte Carlo simulations in Fig.~\ref{fig:simu} for different values of $n, p$ and $q$. Fig.~\ref{fig:simu} also includes the asymptotic approximation derived and discussed in the next section. 
As it can be seen, balls of radius less than one cover a very limited  portion of the space with large dimensions. For example with $(n, p, q)=(8,4,4)$ the radius contraint $r\leq 1$ corresponds to balls covering a maximum of $0.000042 \% $ of the space which corresponds, according to the Gilbert-Varshamov bound,  to code with at least $2.4 \times 10^4$ elements, or equivalently $14.5$ bits.

\section{Asymptotic Analysis}

As shown in the previous section, it is possible, for any radius, to derive exactly the volume of metric balls with specific values of $n,p,q$. However, the resulting expressions are  cumbersome  in large dimensions.  In this section, starting from the reformulation~(\ref{eq:dvre}), the volume is analyzed through asymptotics of the determinant $D_p(\nu)$, providing good and relevant approximations for large-dimensional Grassmann manifolds.  Again, we assume $p\leq q < n $ with $p+ q\leq n$ and other cases can be treated by symmetry. 

\subsection{Asymptotic Volume via Random Matrix Theory}
For reasons that will become clear later, we consider the linear transforms $y_{j}=2x_{j}-1$, $j=1,\dots,p$ in the integral~(\ref{eq:Dp}). After calculating the jacobians associated with the transforms, we have
\begin{eqnarray}
\widetilde{D}_{p}(\nu)&=& \e^{\imag\frac{\nu}{2}p}D_{p}(\nu)\label{eq:DtD} \\
&=&  \tilde{v}_{n,p,q} \int\dots\int_{-1\leq y_{j}\leq1}\Delta^{2}(\ybf)\prod_{j=1}^{p}\left(1-y_{j}\right)^{q-p}\left(1+y_{j}\right)^{n-p-q}\e^{-\imag\frac{\nu}{2}y_{j}}\dd y_{j}, \label{eq:Dpy}
\end{eqnarray}
where
\begin{equation}\label{eq:ct}
\tilde{v}_{n,p,q} =2^{-p(n-p)}v_{n,p,q}.
\end{equation}
One can interpret $\widetilde{D}_{p}(\nu)$ as characteristic function of the random variable
\begin{equation}\label{eq:Y}
Y=\sum_{j=1}^{p}\frac{y_{j}}{2}
\end{equation}
over the so-called Jacobi ensemble 
\begin{equation}\label{eq:Jacobi}
f(\ybf)=\tilde{v}_{n,p,q} \, \Delta^{2}(\ybf)\prod_{j=1}^{p}\left(1-y_{j}\right)^{q-p}\left(1+y_{j}\right)^{n-p-q}.
\end{equation}
In this form, the asymptotic behavior of the linear spectral statistics~(\ref{eq:Y}) is a well-investigated subject in random matrix theory. 
Specifically, by using the result~\cite[Th.~3.2]{Johansson1997} straightforward manipulations\footnote{Namely, with the notations in~\cite{Johansson1997},  $g(x)=-i\frac{\nu}{2}x$ is a linear combination of only the first-order Chebyshev polynomial, then by identifications  $a=q-p$, $b=n-p-q$, and $c_1=-i\frac{\nu}{2}$, Theorem~3.2 from~\cite{Johansson1997} gives $ \log \mathbb{E}\left[\exp(\sum g(y_j))\right] \to \frac{1}{8} c_1^2 - \frac{1}{2}(a-b)c_1$ as $p \to \infty$ and where the expectation is over $f(\ybf)$ in \eqref{eq:Jacobi}.} show that~(\ref{eq:Dpy}) converges to
\begin{eqnarray}
\widetilde{D}_{p}(\nu) &=& \mathbb{E}\left[e^{-i\nu Y}\right]\\
  &\simeq&\e^{-\imag\nu\left(\frac{n-2q}{4}\right)-\frac{\nu^{2}}{32}} \label{eq:DpT}
\end{eqnarray}
in the regime
\begin{equation}\label{eq:rg}
n,p,q\to\infty,~~~\text{with fixed}~q-p~\text{and}~n-p-q.
\end{equation}
To wit, in the asymptotic regime~(\ref{eq:rg}) the random variable~(\ref{eq:Y}) follows a Gaussian distribution with mean and variance read off from~(\ref{eq:DpT}) as
\begin{equation}\label{eq:mevaasy}
\mathbb{E}\left[Y\right]=\frac{n-2q}{4},~~~~\mathbb{V}\left[Y\right]=\frac{1}{16}.
\end{equation}
This is a central limit theorem for the linear statistics~(\ref{eq:Y}) of the Jacobi ensemble~(\ref{eq:Jacobi}). By the relation~(\ref{eq:DtD}), we have
\begin{equation}\label{eq:Dpasy}
D_{p}(\nu) \simeq \e^{-\imag\nu\left(\frac{n+2p-2q}{4}\right)-\frac{\nu^{2}}{32}}.
\end{equation}
Inserting this into~(\ref{eq:dvre}) an asymptotic representation of the volume of metric balls is obtained as
\begin{equation}\label{eq:dvar}
\mu\left(B\left(r\right)\right)\simeq\frac{1}{2\pi}\int_{-\infty}^{\infty}\frac{\imag}{\nu}\left(1-\e^{\imag r^2\nu}\right)\e^{-\imag\nu\left(\frac{n+2p-2q}{4}\right)-\frac{\nu^{2}}{32}}\dd\nu.
\end{equation}
The imaginary part of the integrand is an odd function which integrates to zero. 
The real part is even, from which an asymptotic volume formula is obtained as 
\begin{equation}\label{eq:vasy}
\mu\left(B\left(r\right)\right)\simeq \frac{1}{2}\text{erf}\left(2\sqrt{2}\alpha\right) - \frac{1}{2}\text{erf}\left(2\sqrt{2}\left(\alpha-r^2\right)\right),
\end{equation}
where
\begin{equation}
\alpha=\frac{1}{4}(n+2p-2q)
\end{equation}
and 
\begin{equation}
\text{erf}(x)=\frac{2}{\sqrt{\pi}}\int_{0}^{x}\e^{-t^{2}}\dd t
\end{equation}
is the Gauss error function. 

Although the derived volume formula~(\ref{eq:vasy}) is asymptotically tight in the regime~(\ref{eq:rg}), it may not be very accurate when used as a finite-size approximation.  
The asymptotic~\eqref{eq:DpT} is obtained by letting the size $p$ of the product in the Jacobi ensemble~\eqref{eq:Jacobi} grow to infinity, while  $n,q \to \infty $ keeping the exponents $a = q-p$ and $b=n-p-q$ fixed in~\eqref{eq:Jacobi}. 
As it will be observed below (see Figures~\ref{fig:differentab} and~\ref{fig:Hellinger_distance}), the convergence to the asymptotic distribution is slower when either $|a - b|$, $a$ or $b$ is large, leading to poor approximations with small $p$. 
This fact motivates us to find finite-size corrections to the asymptotic mean and variance~(\ref{eq:mevaasy}) while preserving simplicity of the form~(\ref{eq:vasy}).

\subsection{Finite-size Corrections via Exact Moments}
The idea here is to use exact moments of the linear statistics $Y$ to construct a volume approximation instead of using the asymptotic ones~(\ref{eq:mevaasy}). 
In consistence with the asymptotic Gaussianity in~(\ref{eq:DpT}), we consider a Gaussian approximation of the random variable $Y$ using the first two moments. The exact moments of $Y$ can be recursively obtained via the connection between the moment generating function and an ordinary differential equation. Specifically, the moment generating function of $Y=\sum_{j=1}^{p}y_{j}/2$ is given by 
\begin{equation}\label{eq:MGF}
M_{p}(\nu)=\tilde{v}_{n,p,q}  \int\dots\int_{-1\leq y_{j}\leq1}\Delta^{2}(\ybf)\prod_{j=1}^{p}\left(1-y_{j}\right)^{a}\left(1+y_{j}\right)^{b}\e^{-\frac{\nu}{2}y_{j}}\dd y_{j},
\end{equation}
where
\begin{equation}\label{eq:JEP}
a=q-p,~~~~b=n-p-q
\end{equation}
and $\tilde{v}_{n,p,q} $ is as defined in~(\ref{eq:ct}). By definition, we have
\begin{equation}\label{eq:CGF}
\log M_{p}(\nu)=\sum_{j=1}^{\infty}\kappa_{j}\frac{(-1)^{j}\nu^{j}}{j!},
\end{equation}
where $\kappa_{j}$ denotes the $j$-th cumulant of $Y$. The derivative of cumulant generating function~(\ref{eq:CGF}) satisfies, up to a translation in $\nu$, a nonlinear second-order differential equation~\cite{2010BCE}, which in our notations reads
\begin{eqnarray}\label{eq:ODE}
\left(2\nu\sigma''(\nu)\right)^{2}&=&\left(\sigma(\nu)-\nu\sigma'(\nu)+2(2p+a+b)\sigma'(\nu)\right)^{2}\nonumber\\
& & + 4\left(\sigma(\nu)-\nu\sigma'(\nu)-p(p+b)\right)\left(\left(2\sigma'(\nu)\right)^{2}-2a\sigma'(\nu)\right),
\end{eqnarray}
where
\begin{equation}
\sigma(\nu)=\sum_{j=1}^{\infty}\kappa_{j}\frac{(-1)^{j}\nu^{j}}{2^{j}(j-1)!}-\frac{p\nu}{4}+p(p+b).
\end{equation}
Inserting this into~(\ref{eq:ODE}), the cumulants can be calculated in a recursive manner. The first three cumulants are 
\begin{eqnarray}
\kappa_{1}&=&\frac{p(n-2q)}{2n},\\
\kappa_{2}&=&\frac{pq(n-p)(n-q)}{n^{2}\left(n^{2}-1\right)},\\
\kappa_{3}&=&-\frac{2pq(n-2p)(n-2q)(n-p)(n-q)}{n^{3}\left(n^{4}-5n^{2}+4\right)},
\end{eqnarray}
where we have substituted the parameters according to~(\ref{eq:JEP}). Now we approximate the random variable $Y$ by a Gaussian with mean and variance
\begin{equation}\label{eq:mevae}
\mathbb{E}\left[Y\right]=\kappa_{1},~~~~\mathbb{V}\left[Y\right]=\kappa_{2},
\end{equation}
so that the corresponding moment generating function is approximated by
\begin{equation}
M_{p}(\nu)\approx\e^{-\kappa_{1}\nu+\frac{\kappa_{2}}{2}\nu^{2}}.
\end{equation}
Comparing the moment generating function~(\ref{eq:MGF}) and the characteristic function~(\ref{eq:Dpy}), we have
\begin{equation}
\widetilde{D}_{p}(\nu)\approx\e^{-\imag\kappa_{1}\nu-\frac{\kappa_{2}}{2}\nu^{2}}.
\end{equation}
Following similar steps that led from~(\ref{eq:Dpasy}) to~(\ref{eq:vasy}), we arrive at a finite-size volume approximation 
\begin{equation}\label{eq:ve}
\mu\left(B\left(r\right)\right)\simeq \frac{1}{2}\text{erf}\left(\frac{\beta}{\sqrt{2\kappa_{2}}}\right)-\frac{1}{2}\text{erf}\left(\frac{\beta-r^2}{\sqrt{2\kappa_{2}}}\right),
\end{equation}
where
\begin{equation}
\label{eq:BetaKappa2}
\beta=\frac{p(n-q)}{n},~~~~\kappa_{2}=\frac{pq(n-p)(n-q)}{n^{2}\left(n^{2}-1\right)}.
\end{equation}
Recall that~\eqref{eq:ve} is guaranteed to be asymptotically tight in the regime~(\ref{eq:rg}) according to the asymptotic Gaussianity of $Y$. 
One can also verified with the change of variables $a=q-p$, $b=n-p-q$ and by letting $p \to \infty$ that  the mean and variance in~\eqref{eq:mevae}  are 
\begin{eqnarray}
\label{eq:mean_ab}
\mathbb{E}\left[Y\right]&=&\frac{(b-a )p}{2(a + b + 2 p)} \longrightarrow \frac{(b-a)}{4}= \frac{n-2q}{4} \\
\mathbb{V}\left[Y\right]&=&\frac{p (a + p) (b + p) (a + b + p)}{(a + b + 2 p)^2 ((a + b + 2 p)^2-1 )}\longrightarrow \frac{1}{16}  \label{eq:variance_ab}
\end{eqnarray}
matching~\eqref{eq:mevaasy}  as expected\footnote{It can be as well verified that the third cumulant is asymptotically canceling $\kappa_{3}\to 0$.}. From~\eqref{eq:mean_ab} and~\eqref{eq:variance_ab} , one sees that the larger $a$, $b$, the slower the convergence of the mean and the variance to their asymptotic values would be.

\subsection{Simulations}

\begin{figure}[t]
\centering
\includegraphics[width=\textwidth]{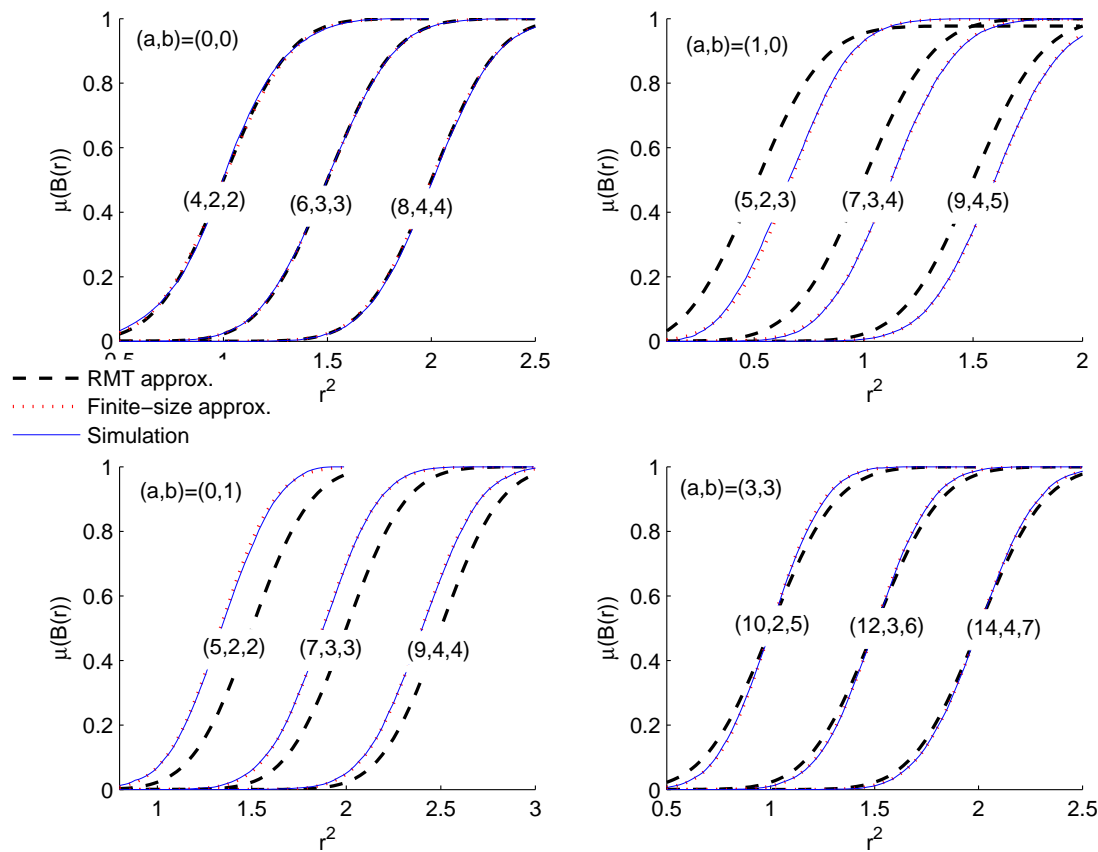}
\vspace{-1.5cm}
\caption{Volumes of metric ball with RMT approx.~(\ref{eq:vasy}) versus finite-size approx.~(\ref{eq:ve}) in the regime~(\ref{eq:rg}). Each graph corresponds to a fixed value of $a=q-p$ and $b=n-q-p$: $(a,b)=(0,0),(1,0),(0,1)$ and $(3,3)$. For every graph, curves from left to right corresponds to  $p=2,3,4$.}\label{fig:differentab}
\end{figure}

\begin{figure}[t]
\centering
\begin{subfigure}{0.5\textwidth}
                \includegraphics[width=1\textwidth]{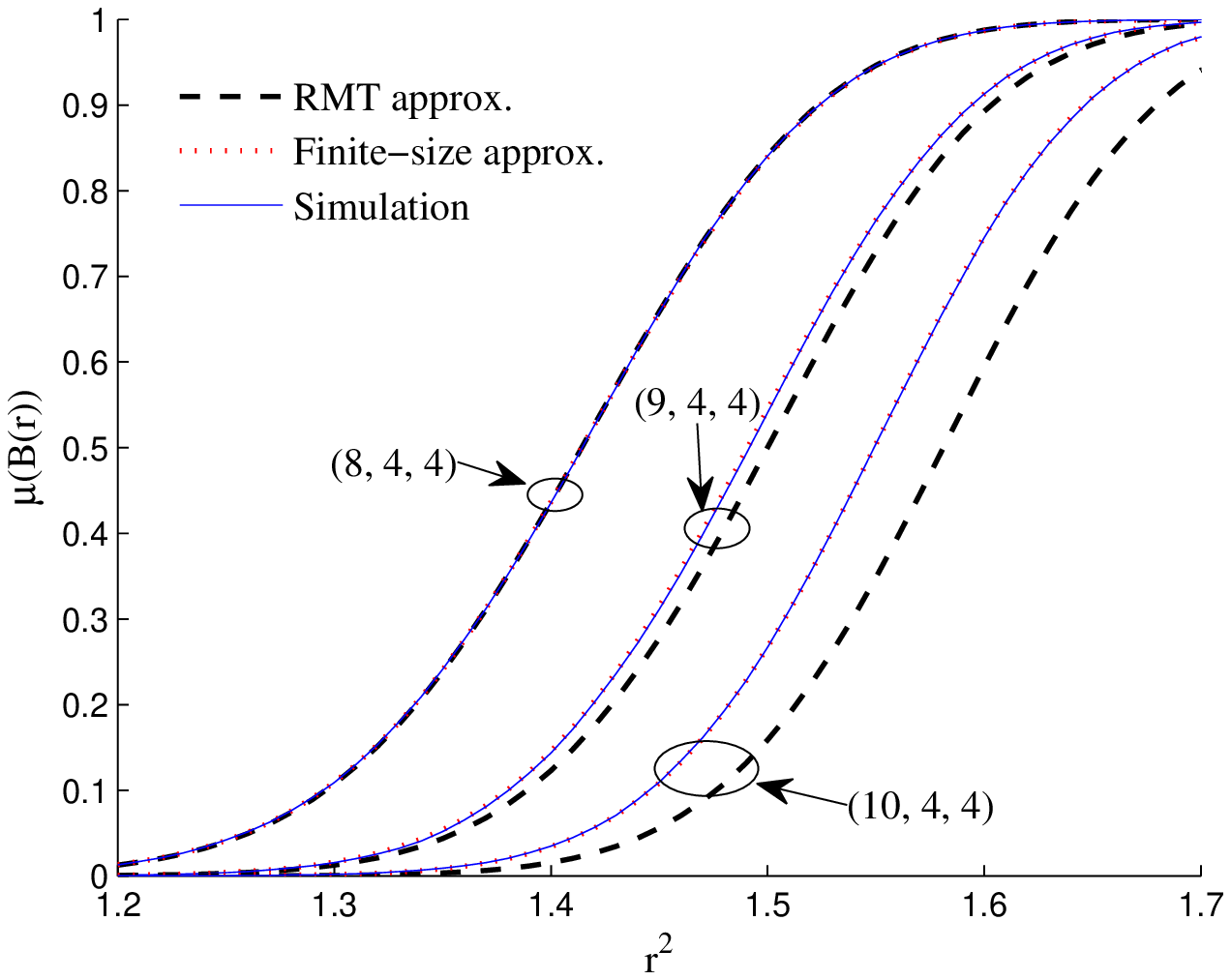}
                \caption{$(n,p,q)=(n,4,4)$}
                \label{fig:p1}
        \end{subfigure}%
\begin{subfigure}{0.5\textwidth}
                \includegraphics[width=1\textwidth]{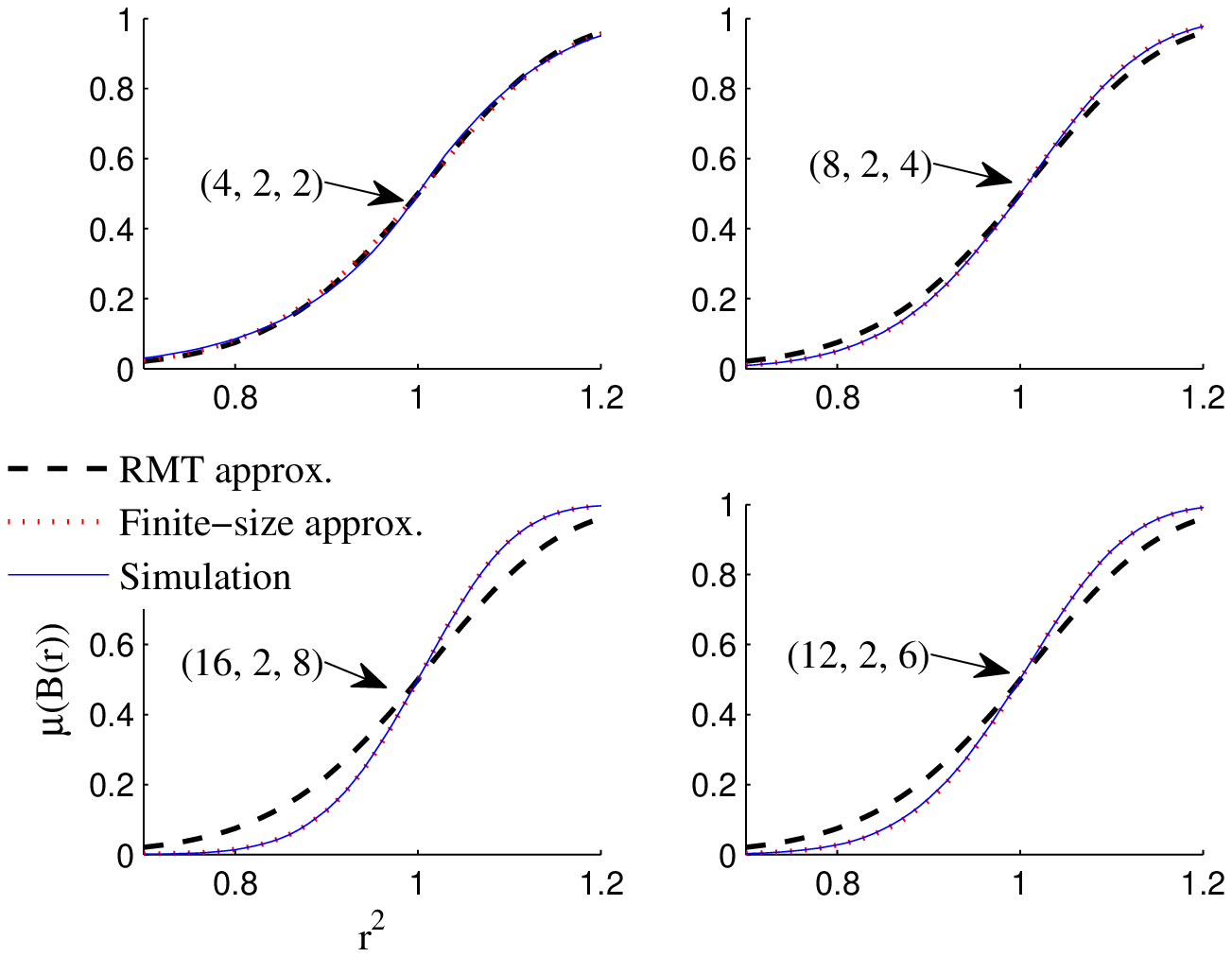}
                \caption{$(n,p,q)=(2q,2,q)$}
                \label{fig:p2}
        \end{subfigure}%
\caption{Volumes of metric ball with RMT approx.~(\ref{eq:vasy}) versus finite-size approx.~(\ref{eq:ve}). Comparison with fixed values of $p$ and growing values of $n,q$.} \label{fig:p12}
\end{figure}

In Figs~\ref{fig:differentab} and~\ref{fig:p12}, we plot the volume of metric ball~(\ref{eq:dv}) calculated by the random matrix theory (RMT) approximation~(\ref{eq:vasy}) as well as the finite-size approximation~(\ref{eq:ve}). As a benchmark, we also provide volume curves by Monte-Carlo simulations. The finite-size correction curves are also included in Fig.~\ref{fig:simu} for comparison with exact expressions. 
In Fig.~\ref{fig:differentab}, asymptotic behavior with the regime~(\ref{eq:rg}) can be observed, i.e. letting $n,p,q$ grow with fixed $q-p$ and $n-p-q$. One verifies that convergence occurs for both approximations, while it can be observed to be much faster for the finite-size correction curves in all cases. The convergence rate of the RMT approximation is dependent of constant values $a=q-p$ and $b=n-p-q$.

\begin{figure}[t]
\centering
\includegraphics[width=0.6\textwidth]{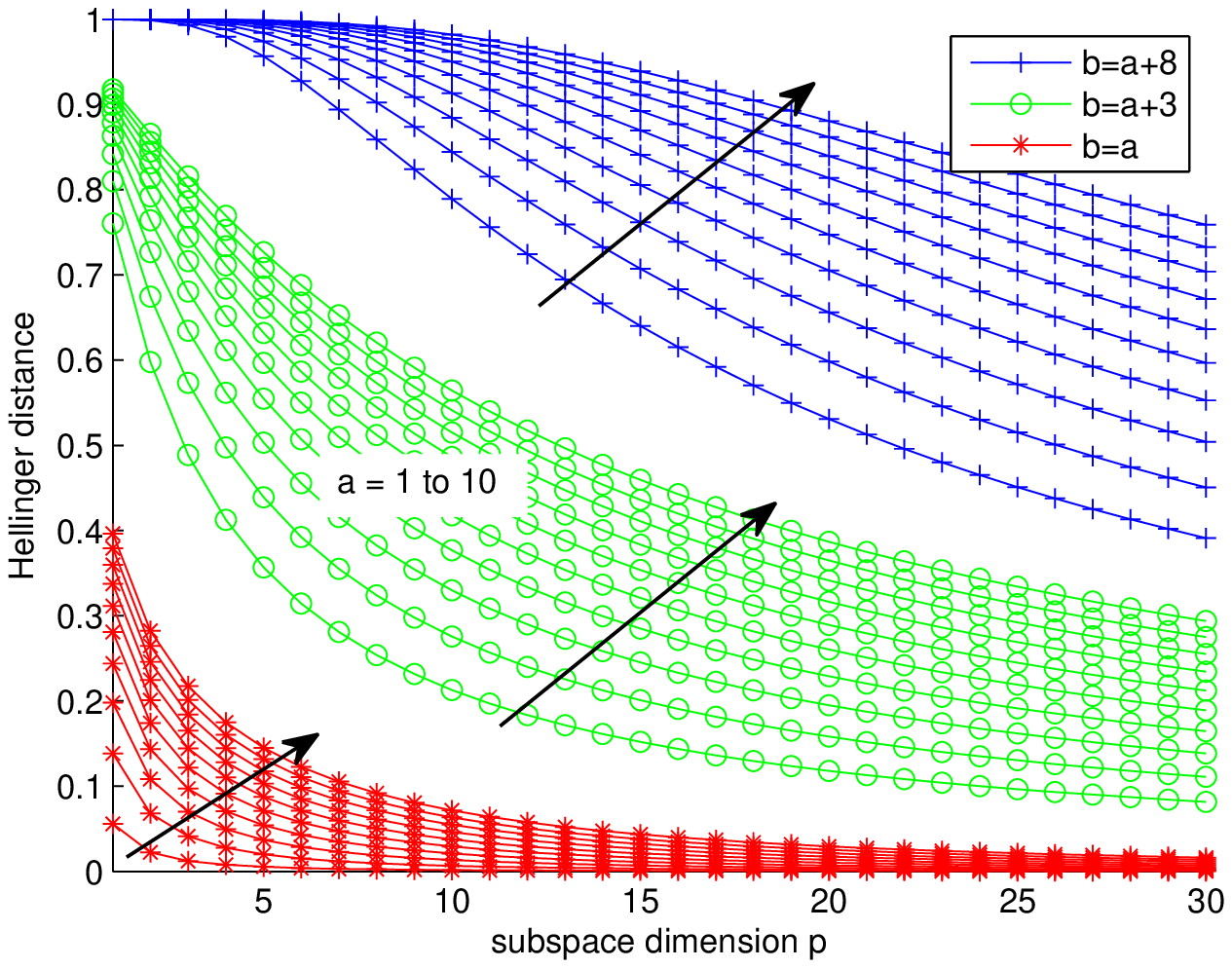}
\vspace{-.5cm}
\caption{Hellinger distance between two Gaussian distributions with moments given in~\eqref{eq:mevaasy} and~\eqref{eq:mevae}, for different fixed values of $a=q-p$ and $b=n-p-q$ as a function of $p$.}\label{fig:Hellinger_distance}
\end{figure}

To evaluate the convergence between the two asymptotic approximations, we compute the divergence between two Gaussian distributions $Y_{1}$ and $Y_{2}$ with means $\mu_{1},\mu_{2} $ and variance $\sigma^2_{1},\ \sigma^2_{2} $ as given in~\eqref{eq:mevaasy} and~\eqref{eq:mevae}, respectively. The Hellinger distance is a type of $f$-divergence which a frequently-used
metric for the spaces of probability distributions~\cite{Liese}. Accordingly, the distance between two Gaussian distributions is~\cite{Kailath} 
\begin{equation}
\label{eq:HellDist}
H(Y_{1},Y_{2}) = \sqrt{1-BC(Y_{1},Y_{2})}
\end{equation}
where
\begin{equation}
BC(Y_{1},Y_{ 2}) = \sqrt{\frac{2\sigma_{1}\sigma_{2}}{\sigma_{1}^2+\sigma_{2}^2}} \, e^{-\frac{1}{4}\frac{(\mu_{1}-\mu_{2})^2}{\sigma_{1}^2+\sigma_{2}^2}}
\end{equation}
is the corresponding Bhattacharyya coefficient. The distance~\eqref{eq:HellDist} is displayed on Fig.~\ref{fig:Hellinger_distance} as a function of $p$ for fixed values of $a$ and $b$. 
For all cases, the  Hellinger distance is converging to zero as expected. A slower rate of convergence is nevertheless verified  for larger $a$, $b$, or  $|a - b|$, and it can be observed that for the maximum considered value $p=30$, the RMT approximation is at more than  half of the maximum distance from finite-size approximation in the highest dimensional cases.

In Fig.~\ref{fig:p12} we consider cases of fixed $p$ and growing values of $n$ and $q$. In Fig.~\ref{fig:p1}  $p=q=4$ and $n=8,9,10$, it is seen that the RMT curves shift horizontally away from the simulated ones as $b=n-2p$ increases. This means a shift in the mean value if we interpret the curves as CDFs of volume density. Simulations indicate that for $p=q$ the RMT approximation~(\ref{eq:vasy}) incurs a nontrivial loss in the mean when the difference $b=n-2p$ is greater than zero. 
In Fig.~\ref{fig:p2}, we consider the case when $n=2q$ with $q=2,4,6$ and $8$ in clock-wise order  for a fixed $p=2$. It is observed that the RMT based curves rotate away from the simulated ones as $a=q-p$ increases. While here $a=b$ and the means of both approximations are equal in~\eqref{eq:mean_ab}, the RMT approximation~(\ref{eq:vasy}) fails to capture the variance of volume density as $q$ increases. For all the cases considered in Fig.~\ref{fig:p12}, the finite-size approximation~(\ref{eq:ve}) matches the simulation almost exactly.

Intuitively, the RMT approximation~(\ref{eq:vasy}) fails to capture the volume curves since the corresponding asymptotic mean and variance~(\ref{eq:mevaasy}) do not involve all the possible parameters $p,q$ and $n$. In particular, the variance~(\ref{eq:mevaasy}) obtained by RMT is a constant. On the contrary, the mean and variance~(\ref{eq:mevae}) used to construct the finite-size approximation~(\ref{eq:ve}) are functions of all the parameters. 

\section{Application to Source Coding on Grassmann Manifolds} 

The asymptotic volume of a ball with its finite size-correction can be applied to evaluate the rate-distortion trade-off of a source quantization in large-dimensional Grassmann manifolds. 

Given a code $\Ccal = \{ C_1,\ldots, C_N \} \subset \grass{C}{n}{p}$ with size $N= |\Ccal|$, consider a uniformly distributed  source on $\grass{C}{n}{q}$ quantized to $\Ccal$ using the chordal distance $d_c$ as a quantization map:
\begin{eqnarray} 
\grass{C}{n}{q} &\to& \Ccal \\
Q  &\mapsto& \arg \min_{C_k \in  \Ccal} d_c(Q,C_k)  .
\end{eqnarray}

A source code is considered optimal if it minimizes the average distortion of the process, i.e. the average square quantization error
\begin{eqnarray}
D(\Ccal) &=& \Ebb_{Q \in \grass{C}{n}{q}}[  \min_k d_c^2(C_k,Q) ]\\
       &=& \int_0^p z dF_{\Ccal}(z)\\ 
			&=&  \int_0^p (1-F_{\Ccal}(z)) dz,
\end{eqnarray}
where $F_{\Ccal}(z) = {\rm Pr} \left\{Q \;|\;  \min_k d_c^2(C_k,Q)\leq  z \right\}$ is the CDF of quantization error. 
The distortion-rate function is the infimum of all possible distortions for a given codesize,  
\begin{equation} D(N) = \inf_{|\Ccal|=N} D(\Ccal). \end{equation}

The quantization map leads to the partition of the Grassmann manifold into  Voronoi cells centered around the codewords which are defined as 
\begin{equation}
V_k= \{ Q \ \;|\;     d_c^2(C_k,Q)\leq  d_c^2(C_j,Q),\, \forall j  \} .
\end{equation}
It follows that the probability of quantization error being less or equal to a value $z$ is given by
\begin{eqnarray}
 F_{\Ccal}(z) &=& {\rm Pr} \left\{Q \;|\;  \min_k d_c^2(C_k,Q)\leq  z \right\} \\
 &=& {\rm Pr} \left\{  \cup_{k=1}^{N} \left\{Q\in (B_{C_k}(\sqrt{z}) \cap V_k) \right\} \right\} .
\end{eqnarray}
As points on the manifold belonging to the Voronoi cells' borders appear with probability zero, we further have
\begin{eqnarray}
 F_{\Ccal}(z) &=& \sum_{k=1}^N {\rm Pr} \left\{  B_{C_k}(\sqrt{z}) \cap V_k \right\} \\
 &=&\sum_{k=1}^N \mu \left(B_{C_k}(\sqrt{z}) \cap V_k \right). \label{eq:Fc(z)}
\end{eqnarray}

The CDF of quantization errors~\eqref{eq:Fc(z)} is actually equal to  the volume of balls until some border effect, i.e.  $F_{\Ccal}(z)  = N \mu(B(\sqrt{z})) $ on the interval $[0,\, \varrho^2]$ where $\varrho$ is the \emph{kissing radius} of the code~\cite{ISIT11}: the shortest  distance from a codeword to the border of a Voronoi cell.
In general, for any $k$, $\left(B_{C_k}(\sqrt{z}) \cap V_k \right) \subseteq B_{C_k}(\sqrt{z}) $, and so for any $z$ one has  $F_{\Ccal}(z) \leq N \mu(B(\sqrt{z}))$.

As shown in~\cite{Dai08}, by using this upper bound to design an ideal distribution  $F^*_{\Ccal}(z)= N \mu(B(\sqrt{z})) $ on  $[0,\, z^*]$ such that $ N \mu(B(\sqrt{z^*}))=1$ leads to a lower bound on distortions, i.e.  $D(N) \geq \int_0^{ z^*} z dF^*_{\Ccal}(z) $. 
Following this principle, after estimating $z^*$ by inverting the derived volume~\eqref{eq:ve} from the previous section and by direct integration we obtain the following approximation to the rate-distortion trade-off
\begin{equation}
D(N)  \gtrsim \beta -N \sqrt{ \frac{ \kappa_2 }{2 \pi }}\left( e^{-\left(\text{erf}^{-1}\left(\text{erf}\left(\frac{\beta }{\sqrt{2 \kappa_2 }}\right)-\frac{2}{N}\right)\right)^2} - e^{-\frac{\beta^2}{2 \kappa_2 }}\right), 
\label{eq:distortionLB}
\end{equation}
which is asymptotically a lower bound for large-dimensional codes as $ n,p,q\to\infty$ with fixed $q-p$ and $n-p-q$. 
The parameters $\beta$ and $\kappa_2$ are given in~\eqref{eq:BetaKappa2}. 

\begin{figure}[!t]
\centering
\includegraphics[width=0.8\textwidth]{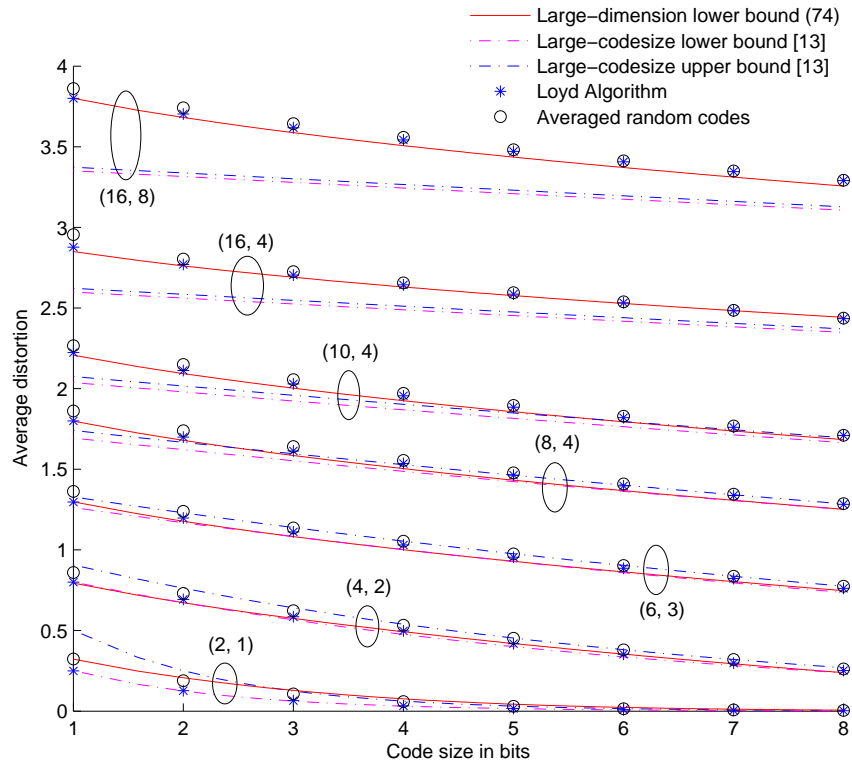}
\caption{The large dimension bound~\eqref{eq:distortionLB} compared to simulated average distortions and high-resolution bounds~\cite{Dai08}.}
\label{fig:distortion}
\end{figure}

The approximation~\eqref{eq:distortionLB} is compared to simulations in Fig.~\ref{fig:distortion} for different values of $n$, $p$, and with $q=p$. Codes with cardinality between $2$ and $256$, i.e., between one and eight bits, are considered. Rate-distortion trade-offs have been numerically minimized by applying vector quantization based on Lloyd's algorithm. In addition to the numerical vector quantization results,  averaged distortions of random codes are shown, which by construction  provide an upper bound on $D(N)$. It is visible that the derived approximation~\eqref{eq:distortionLB} is asymptotically a lower bound in large dimensions, while obviously not for the smallest case $(n,p)=(2,1)$. For all other dimensions, it provides a rather good approximation for every cardinality.  

The large-dimension bound~\eqref{eq:distortionLB} is further compared to the high-resolution bounds in~\cite[Th. 2]{Dai08}. The results in~\cite{Dai08} are asymptotics in a different regime with a range of validity given a  sufficiently large codesize. A necessary condition can be found in~\cite{Dai08}. Here, the lower bound for code size where the results~\cite{Dai08} start to apply would be  $0$, $1$, $5.4$, $14.6$, $27.1$, $68.6$ and $114.1$ bits for the cases $(n,p)=(2, 1)$, $(4, 2)$, $(6, 3)$, $(8, 4)$,  $(10, 4)$, $(16, 4)$ and $(16, 8)$, respectively. These are plotted in Fig.~\eqref{eq:distortionLB} from bottom up.The high-resolution bounds in~\cite{Dai08} provides also very good approximations  of the rate-distortion trade-off for much smaller codesizes, outside of their given range of validity. Nevertheless, in the large-dimensional regime and with fixed code cardinality, one can observe in Fig.~\ref{fig:distortion}  a trend in the slope of the rate-distortion trade-off which is not captured by the high-resolution bounds. The bound~\eqref{eq:distortionLB} provides a good approximation in almost all cases. This is the consequence of the volume evaluation~\eqref{eq:ve} for any radius.  Moreover, the bounds in~\cite{Dai08} depends on a fastly-decreasing constant $c_{n,p,q} \to 0$ as $n,p,q \to \infty$ which might lead to numerical computation issues in the large-dimension regime. For example, its value is $c_{n,p,q}=4.2 \times 10^{-5}$ for $(n,p)=(8,4)$  and $c_{n,p,q}=4.5 \times 10^{-35}$ for $(n,p)=(16,8)$, while we faced numerical computation errors for $(n,p)=(32,16)$ due to machine precision.

\section{Conclusion}

We evaluated the volume of a metric ball in Grassmann manifolds. The case of a center with mismatched dimension is considered and accordingly we discuss generalizations of
the chordal distance to subspaces with unequal dimensions. First, a new symmetry property of the volume of a metric ball is presented. 
Then, multivariate integration of the volume of a ball with any radius is performed. We reduce the multivariate integration problem to a single-fold integral related to
Fourier transform. We also provide explicit examples in small dimensions for any radius from the obtained formula. 
For large dimensions, the derived integral expression provides a tractable starting point for asymptotic analysis. From the asymptotic behavior of the time-dependent Jacobi ensemble and by moment-matching techniques, we provide a simple asymptotic volume formula which provides a tight approximation in finite-size dimensions. 
This allows us to precisely quantify the rate-distortion trade-off of source coding problems in large-dimensional Grassmann manifolds.

The results presented in this paper are valid for the Grassman manifold over the complex field. Using the same methodology for a generalization to the real field does not appear trivial. The exact volume formulas were explicitly derived via the Andr\'{e}ief identity. 
The Andr\'{e}ief identity is also intrinsically connected with the asymptotic analysis presented here. By using this identity, one is able to relate the problem to a Hankel or Toeplitz determinant whose asymptotical behaviors have been extensively studied in statistics. 
In order to proceed with Andr\'{e}ief identity, the square of the Vandermonde determinant in the volume element is instrumental. 
In contrast, for the real Grassmann manifolds, the Vandermonde determinant in the volume element is not squared and one cannot thus directly reduce the problem in a similar fashion.
It remains thus an open problem for future research to identify comparable accurate approximation techniques for the real case.


\appendices


\section{Proof of Symmetry Relationship~\eqref{eq:sym3}}
\label{app:sym3}
Given $P \in \grass{C}{n}{p}$  with orthogonal complement $P^\bot    \in \grass{C}{n}{n-p}$, 
and similarly given $Q \in \grass{C}{n}{q}$ with orthogonal complement $Q^\bot \in \grass{C}{n}{n-q}$, one has
 \begin{eqnarray}
p  &=& \| \Pbf^\dag \Qbf \|_F^2 + \| \Pbf^\dag \Qbf^\bot \|_F^2,
\end{eqnarray} 
which leads to 
 \begin{eqnarray}
d_c^2(P,Q) &=& p - d_c^2(P,Q^\bot).  
\end{eqnarray}

Given a point $Q\in \grass{C}{n}{q}$ such that $d_c(P,Q) \geq r$, it follows that $d_c(P^\bot,Q) \leq \sqrt{p-r^2}$, and thus $Q \notin B_{\scriptscriptstyle P,q}(r)$ implies  $Q \in B_{\scriptscriptstyle P^\bot,q}(\sqrt{p-r^2})$.  Reciprocally $Q \in B_{\scriptscriptstyle P,q}(r)$ implies  $Q \notin B_{\scriptscriptstyle P^\bot,q}(\sqrt{p-r^2})$, so that 
\begin{eqnarray}
B_{\scriptscriptstyle P,q}(r) \cap B_{\scriptscriptstyle P^\bot,q}(\sqrt{p-r^2}) &=&  \emptyset \\
B_{\scriptscriptstyle P,q}(r) \cup B_{\scriptscriptstyle P^\bot,q}(\sqrt{p-r^2}) &=&  \grass{C}{n}{q}.
\end{eqnarray}
Finally, 
\begin{equation} \mu(B_{\scriptscriptstyle P,q}(r))+ \mu(B_{\scriptscriptstyle P^\bot,q}(\sqrt{p-r^2}))=1, 
\end{equation}  
and using \eqref{eq:sym1}, \eqref{eq:sym2}, we obtain~\eqref{eq:sym3}.

\section{Andr\'{e}ief integral~\cite{1883Andreief}}\label{a:Andreief}
For two $n\times n$ matrices $\mathbf{A}(\mathbf{x})$ and $\mathbf{B}(\mathbf{x})$, with the respective $ij$-th entry being functions $A_{i}(x_{j})$ and $B_{i}(x_{j})$, and a function $f(\cdot)$ such that the integral $\int_{a}^{b}A_{i}(x)B_{j}(x)f(x)\dd x$ exists, the multiple integral of the product of the determinants can be evaluated as
\begin{equation}\label{eq:AI}
\int\dots\int_{\mathcal{D}}\det\big(\mathbf{A}(\mathbf{x})\big)\det\big(\mathbf{B}(\mathbf{x})\big)\prod_{i=1}^{n}f(x_{i})\mathrm{d}x_{i}=  \det\left(\int_{a}^{b}A_{i}(x)B_{j}(x)f(x)\dd x\right),
\end{equation}
where $\mathcal{D}=\{a\leq x_{n}\leq\ldots\leq x_{1}\leq b\}$.

\section{Exemples of Exact Volume Computation}\label{a:exemples}
We give explicit expression of $\mu\left(B\left(r\right)\right)$ obtained by~\eqref{eq:main} and~\eqref{eq:qpn}. 
The volumes have different polynomial representations between different consecutive integer values of $r^2$, i.e. on the intervals $[0,\,1]$, $[1,\,2],\ldots$, $[p-1,\,p]$. The expressions given here are valid for any $r^2\in [0,\, p]$. 
It can be verified for $r\leq1$ that the expressions simplify (e.g. to a monomial for $p=q$) and match the results in~\cite{Dai08}. 
\subsubsection{$n=4$ and $p=q=2$}
\begin{eqnarray*}
\mu\left(B\left(r\right)\right)&=&-\frac{7}{2}+8 r^2-6 r^4+2 r^6-   \frac{\left(r^2-1\right)^3 \left(7-2 r^2+r^4\right)}{2 |r^2-1|} 
\end{eqnarray*}

\subsubsection{$n=5$ and $p=q=2$}

\begin{eqnarray*}
\mu\left(B\left(r\right)\right)&=& {\textstyle \frac{17}{2}-\frac{144 r^2}{5}+36 r^4-20 r^6 +\frac{9 r^8}{2} -\frac{\left(r^2-1\right)^4 \left(85-33 r^2+6 r^4+2 r^6\right)}{10 |r^2-1|} } 
\end{eqnarray*}

\subsubsection{$n=5$, $p=2$ and $q=3$}

\begin{eqnarray*}
\mu\left(B\left(r\right)\right)&=&  {\textstyle -\frac{59}{10}+\frac{96 r^2}{5}-24 r^4+16 r^6-\frac{9 r^8}{2} +\left(\frac{59}{10}-\frac{3 r^2}{2}+\frac{9 r^4}{5}-\frac{r^6}{5}\right) |r^2-1|^3}
\end{eqnarray*}

\subsubsection{$n=6$ and $p=q=2$}

\begin{eqnarray*}
\mu\left(B\left(r\right)\right)&=&  {\textstyle -\frac{31}{2}+\frac{480 r^2}{7}-120 r^4+104 r^6-45 r^8+8 r^{10}  -\frac{(r^2-1)^5 \left(217-92 r^2+10 r^4+4 r^6+r^8\right)}{14 |r^2-1|}     }
\end{eqnarray*}

\subsubsection{$n=6$, $p=2$ and $q=3$}

\begin{eqnarray*}
\mu\left(B\left(r\right)\right)&=&  {\textstyle \frac{263}{14}-\frac{576 r^2}{7}+144 r^4-128 r^6+60 r^8-12 r^{10}-\frac{\left(r^2-1\right)^5 \left(-263+100 r^2-38 r^4-12 r^6+3 r^8\right)}{14 |r^2-1|}  }
\end{eqnarray*}

\subsubsection{$n=6$ and $p=q=3$}
\begin{multline*}
\mu\left(B\left(r\right)\right)={\textstyle  -\frac{6547}{28}+\frac{19683 r^2}{28}-\frac{6561 r^4}{7}+729 r^6-\frac{729 r^8}{2}  +\frac{243 r^{10}}{2}-27 r^{12}+\frac{27 r^{14}}{7}-\frac{9 r^{16}}{28} +\frac{r^{18}}{42} } \\
\quad \quad {\textstyle +\frac{6 \left(r^2-1\right)^7-9 |r^2-1|^6-\frac{18}{7}|r^2-1|^8-\frac{1}{28} |r^2-1|^{10}}{|r^2-1|}  +\frac{6 \left(r^2-2\right)^7+9 |r^2-2|^6+\frac{18}{7}|r^2-2|^8+\frac{1}{28}|r^2-2|^{10}}{|r^2-2|} } 
\end{multline*}

\section*{Acknowledgment}
R.-A. Pitaval was supported by the Academy of Finland (Grants 276031, 282938, 283262) and the European Science
Foundation under the COST Action IC1104. L. Wei is supported by the Academy of Finland (Grant 251170) and Nokia Foundation. O. Tirkkonen is supported by the Academy of Finland (Grant 284725).
J. Corander is supported by the Academy of Finland (Grant 251170). 
\addcontentsline{toc}{chapter}{Bibliography}
\bibliographystyle{IEEEtran}
\bibliography{ISIT15}

\begin{thebibliography}{10}
\providecommand{\url}[1]{#1}
\csname url@samestyle\endcsname
\providecommand{\newblock}{\relax}
\providecommand{\bibinfo}[2]{#2}
\providecommand{\BIBentrySTDinterwordspacing}{\spaceskip=0pt\relax}
\providecommand{\BIBentryALTinterwordstretchfactor}{4}
\providecommand{\BIBentryALTinterwordspacing}{\spaceskip=\fontdimen2\font plus
\BIBentryALTinterwordstretchfactor\fontdimen3\font minus
  \fontdimen4\font\relax}
\providecommand{\BIBforeignlanguage}[2]{{%
\expandafter\ifx\csname l@#1\endcsname\relax
\typeout{** WARNING: IEEEtran.bst: No hyphenation pattern has been}%
\typeout{** loaded for the language `#1'. Using the pattern for}%
\typeout{** the default language instead.}%
\else
\language=\csname l@#1\endcsname
\fi
#2}}
\providecommand{\BIBdecl}{\relax}
\BIBdecl

\bibitem{Conway1996}
J.~H. Conway, R.~H. Hardin, and N.~J.~A. Sloane, ``Packing lines, planes, etc.:
  Packings in {G}rassmannian space,'' \emph{Exper. Math.}, vol.~5, pp.
  139--159, 1996.

\bibitem{Edelman}
A.~Edelman, T.~Arias, and S.~T. Smith, ``The geometry of algorithms with
  orthogonality constraints,'' \emph{{SIAM} J. Matrix Anal. Appl.}, vol.~20,
  no.~2, pp. 303--353, 1998.

\bibitem{YeLim}
\BIBentryALTinterwordspacing
K.~Ye and L.-H. Lim, ``Distance between subspaces of different dimensions,''
  2014. [Online]. Available: \url{http://arxiv.org/abs/1407.0900}
\BIBentrySTDinterwordspacing

\bibitem{AgrawalTIT01}
D.~Agrawal, T.~Richardson, and R.~Urbanke, ``Multiple-antenna signal
  constellations for fading channels,'' \emph{IEEE Trans. Inf. Theory},
  vol.~47, no.~6, pp. 2618--2626, Sep. 2001.

\bibitem{Tse}
L.~Zheng and D.~Tse, ``Communication on the {G}rassmann manifold: a geometric
  approach to the noncoherent multiple-antenna channel,'' \emph{IEEE Trans.
  Inf. Theory}, vol.~48, no.~2, pp. 359 --383, Feb. 2002.

\bibitem{AshikhminTIT2010}
A.~Ashikhmin and A.~Calderbank, ``Grassmannian packings from operator
  {Reed-Muller} codes,'' \emph{IEEE Trans. Inf. Theory}, vol.~56, no.~11, pp.
  5689--5714, Nov. 2010.

\bibitem{LoveTIT03}
D.~J. Love, R.~W. Heath~Jr., and T.~Strohmer, ``Grassmannian beamforming for
  multiple-input multiple-output wireless systems,'' \emph{IEEE Trans. Inf.
  Theory}, vol.~49, no.~10, pp. 2735--2747, Oct. 2003.

\bibitem{Mukkavilli03}
K.~Mukkavilli, A.~Sabharwal, E.~Erkip, and B.~Aazhang, ``On beamforming with
  finite rate feedback in multiple-antenna systems,'' \emph{IEEE Trans. on Inf.
  Theory}, vol.~49, no.~10, pp. 2562--2579, Oct. 2003.

\bibitem{Barg}
A.~Barg and D.~Nogin, ``Bounds on packings of spheres in the {G}rassmann
  manifold,'' \emph{IEEE Trans. Inf. Theory}, vol.~48, no.~9, pp. 2450--2454,
  Sep. 2002.

\bibitem{henkel}
O.~Henkel, ``Sphere packing bounds in the {G}rassmann and {S}tiefel
  manifolds,'' \emph{IEEE Trans. Inf. Theory}, vol.~51, no.~10, pp. 3445--3456,
  Oct. 2005.

\bibitem{Roh}
J.~Roh and B.~Rao, ``Transmit beamforming in multiple-antenna systems with
  finite rate feedback: a {VQ}-based approach,'' \emph{IEEE Trans. Inf.
  Theory}, vol.~52, no.~3, pp. 1101--1112, Mar. 2006.

\bibitem{Gian06}
P.~Xia and G.~Giannakis, ``Design and analysis of transmit-beamforming based on
  limited-rate feedback,'' \emph{IEEE Trans. on Signal Process.}, vol.~54,
  no.~5, pp. 1853--1863, May 2006.

\bibitem{Dai08}
W.~Dai, Y.~Liu, and B.~Rider, ``Quantization bounds on {G}rassmann manifolds
  and applications to {MIMO} communications,'' \emph{IEEE Trans. Inf. Theory},
  vol.~54, no.~3, pp. 1108--1123, Mar. 2008.

\bibitem{DaiOnOff}
W.~Dai, Y.~Liu, B.~Rider, and V.~Lau, ``On the information rate of {MIMO}
  systems with finite rate channel state feedback using beamforming and power
  on/off strategy,'' \emph{IEEE Trans. Inf. Theory}, vol.~55, no.~11, pp.
  5032--5047, Nov. 2009.

\bibitem{DaiGrowth07}
W.~Dai, B.~Rider, and Y.~Liu, ``Volume growth and general rate quantization on
  {Grassmann} manifolds,'' in \emph{Proc. IEEE Global Telecom. Conf.}, Nov.
  2007, pp. 1441--1445.

\bibitem{2014Zhu}
\BIBentryALTinterwordspacing
D.~Zhu, B.~Li, and P.~Liang, ``Normalized volume of hyperball in complex
  {Grassmann} manifold and its application in large-scale {MU-MIMO}
  communication systems,'' 2014. [Online]. Available:
  \url{http://arxiv.org/abs/1402.4543}
\BIBentrySTDinterwordspacing

\bibitem{JindalISIT06}
N.~Jindal, ``A feedback reduction technique for {MIMO} broadcast channels,'' in
  \emph{Proc. IEEE Int. Symp. Inf. Theory}, July 2006, pp. 2699--2703.

\bibitem{Thomas2014}
\BIBentryALTinterwordspacing
B.~S. Thomas, L.~Lin, L.-H. Lim, and S.~Mukherjee, ``Learning subspaces of
  different dimensions,'' 2014. [Online]. Available:
  \url{http://arxiv.org/abs/1404.6841}
\BIBentrySTDinterwordspacing

\bibitem{LeeTSP14}
J.~H. Lee and W.~Choi, ``Multiuser diversity for secrecy communications using
  opportunistic jammer selection: Secure {DoF} and jammer scaling law,''
  \emph{IEEE Trans. Signal Process.}, vol.~62, no.~4, pp. 828--839, Feb. 2014.

\bibitem{James}
A.~T. James, ``Normal multivariate analysis and the orthogonal group,''
  \emph{Ann. Math. Statist.}, vol.~25, no.~1, pp. 40--75, 1954.

\bibitem{Adler}
M.~Adler and P.~V. Moerbeke, ``Integrals over {G}rassmannians and random
  permutations,'' \emph{Adv. Math.}, vol. 181, no.~1, p. 190–249, Feb. 2004.

\bibitem{Johansson1997}
K.~Johansson, ``On random matrices from the compact classical groups,''
  \emph{Ann. Math.}, vol. 145, no.~3, pp. 519--545, May 1997.

\bibitem{Love:limited}
D.~Love and R.~Heath, ``Limited feedback unitary precoding for spatial
  multiplexing systems,'' \emph{IEEE Trans. Inf. Theory}, vol.~51, no.~8, pp.
  2967--2976, Aug. 2005.

\bibitem{Bachoc08}
C.~Bachoc, Y.~Ben-Haim, and S.~Litsyn, ``Bounds for codes in products of
  spaces, {G}rassmann, and {S}tiefel manifolds,'' \emph{IEEE Trans. Inf.
  Theory}, vol.~54, no.~3, pp. 1024 --1035, Mar. 2008.

\bibitem{Wang52}
H.-C. Wang, ``Two-point homogeneous spaces,'' \emph{Ann. Math}, vol.~55, no.~1,
  pp. 177--191, Jan. 1952.

\bibitem{PitavalITW13}
R.-A. Pitaval and O.~Tirkkonen, ``Flag orbit codes and their expansion to
  {S}tiefel codes,'' in \emph{Proc. IEEE Inf. Theory Workshop}, Sep. 2013, pp.
  1--5.

\bibitem{PitavalAsilomarBall}
------, ``Volume of ball and {H}amming-type bounds for {S}tiefel manifold with
  {E}uclidean distance,'' in \emph{Proc. Asilomar Conf. on Sig., Syst. and
  Comp.}, Nov. 2012, pp. 483--487.

\bibitem{Han}
G.~Han and J.~Rosenthal, ``Unitary space-time constellation analysis: An upper
  bound for the diversity,'' \emph{IEEE Trans. Inf. Theory}, vol.~52, no.~10,
  pp. 4713--4721, Oct. 2006.

\bibitem{2010BCE}
E.~Basor, Y.~Chen, and T.~Ehrhardt, ``Painlev\'{e} {V} and time-dependent
  {Jacobi} polynomials,'' \emph{J. Phys. A: Math. Theor.}, vol.~43, 2010.

\bibitem{1883Andreief}
C.~Andr\'{e}ief, ``Note sur une relation entre les int\'{e}grales d\'{e}finies
  des produits des fonctions,'' \emph{M\'{e}m. de la Soc. Sci. Bordeaux 2},
  1883.

\bibitem{Chiani}
M.~Chiani, M.~Z. Win, and A.~Zanella, ``On the capacity of spatially correlated
  {MIMO} {Rayleigh}-fading channels,'' \emph{IEEE Trans. Inf. Theory}, vol.~49,
  no.~10, pp. 2363--2371, Oct. 2003.

\bibitem{2007GR}
I.~S. Gradshteyn and I.~M. Ryzhik, \emph{Table of Integrals, Series, and
  Products.}\hskip 1em plus 0.5em minus 0.4em\relax $7$th Ed., San Diego:
  Academic Press, 2007.

\bibitem{Liese}
F.~Liese and I.~Vajda, ``On divergences and informations in statistics and
  information theory,'' \emph{IEEE Trans. Inf. Theory}, vol.~52, no.~10, pp.
  4394--4412, Oct. 2006.

\bibitem{Kailath}
T.~Kailath, ``The divergence and {Bhattacharyya} distance measures in signal
  selection,'' \emph{IEEE Trans. Commun. Tech.}, vol.~15, no.~1, pp. 52--60,
  Feb. 1967.

\bibitem{ISIT11}
R.-A. Pitaval, O.~Tirkkonen, and S.~D. Blostein, ``Density and bounds for
  {Grassmannian} codes with chordal distance,'' in \emph{Proc. IEEE Int. Symp.
  Inf. Theory}, Aug. 2011, pp. 1--5.

\end{thebibliography}

\end{document}